\documentclass[12pt]{article}
\title{Higher $\eta_c(nS)$ and $\eta_b (nS)$ mesons}
\author{A.M. Badalian$^a$\footnote{E-mail:badalian@itep.ru}
 and B.L.G. Bakker$^b$\\
\\
$^a$ Institute of Theoretical and Experimental Physics,\\
117218, Moscow, B.Cheremushkinskaya, 25, Russia \\
$^b$ Department of Physics and Astronomy,\\
Vrije Universiteit, Amsterdam, The Netherlands}
\date{\today}
\begin{document}
\maketitle
\def\la{\mathrel{\mathpalette\fun <}}
\def\ga{\mathrel{\mathpalette\fun >}}
\def\fun#1#2{\lower3.6pt\vbox{\baselineskip0pt\lineskip.9pt
\ialign{$\mathsurround=0pt#1\hfil ##\hfil$\crcr#2\crcr\sim\crcr}}}
\newcommand{\vep}{\mbox{\boldmath$p$}}
\newcommand{\ver}{\mbox{\boldmath$r$}}
\newcommand{\ves}{\mbox{\boldmath$s$}}
\newcommand{\lan}{\langle}
\begin{abstract}

\noindent
The hyperfine splittings in heavy quarkonia are studied in a
model-independent way using the experimental data on di-electron
widths.  Relativistic correlations are taken into account together with
the smearing of the spin-spin interaction. The radius of smearing is
fixed by the known  $J/\psi-\eta_c(1S)$ and $\psi(2S)-\eta'_c(2S)$
splittings and appears to be small, $r_{ss} \cong 0.06$ fm.
Nevertheless, even with such a small radius an essential suppression of
the hyperfine splittings ($\sim 50\%)$  is observed in bottomonium. For
the $nS~ b\bar b$ states $(n=1,2,\dots,6)$ we predict the values (in
MeV) 28, 12, 10, 6, 6, and 3, respectively.  For the $3S$ and $4S$
charmonium states  the splittings 16(2) MeV and 12(4) MeV are
obtained.
\end{abstract}

\section{Introduction}
At present two spin-singlet $S$-wave states $\eta_c (1S)$ and
$\eta_c(2S)$ have been discovered \cite{ref.01}-\cite{ref.03}. Still, no
spin-singlet $\eta_b(nS)$ levels have been seen \cite{ref.04}.
Theoretically the masses of the $\eta_b(nS)$ were predicted in
many papers \cite{ref.05}-\cite{ref.11}, however, the calculated hyperfine
(HF) splittings,
\begin{equation}
 \Delta_{\rm HF} (nS) = M(n^3 S_1) - M(n^1S_0),
\label{1}
\end{equation}
vary in a wide range: from 35 MeV up to 100 MeV  for the $b\bar b~ 1S$
state and for the 2S state between 19 MeV  and 44 MeV \cite{ref.11}.

However, modern theoretical treatments and current experiments taken
together, have produced well-established limits on the factors which
determine the spin-spin potential $V_{\rm HF}(r)$ in heavy quarkonia.
First of all, the wave function (w.f.) at the origin for a given
$n^3S_1(c\bar c $ or $b\bar b$) state can be extracted from the
di-electron width which, as well as some ratios of leptonic widths, are
now measured with high accuracy \cite{ref.12,ref.13}. The quark masses,
the pole (current) mass, present in the correct relativistic approach,
and the constituent mass, used in the nonrelativistic or in more refined
approximations, is also known with good accuracy \cite{ref.14,ref.15}.
Therefore the only uncertainties comes from two sources.

First, in perturbative QCD a strict prescription as to how to choose
the renormalization scale $\mu$ in the strong coupling $\alpha_{\rm
HF}$, which enters $V_{\rm HF}(r)$, is not well established for a bound
state, especially for higher excitations.

Secondly, a smearing of the spin-spin interaction is considered to be
due to relativistic effects \cite{ref.10} but the true size of the
smearing radius $r_{ss}$ is still not fixed.

In our calculations the radius $r_{ss}$ is taken to fit the
$J/\psi-\eta_c(1S)$ and $\psi(2S)-\eta_c(2S)$ splittings. We show that
to reach agreement with experiment the smearing radius should be
$r_{ss} \leq 0.06$ fm. Our value $r_{ss}=0.057$ fm practically
coincides with the number used in Ref. \cite{ref.10}.  However, in
spite of this coincidence the splitting $\Delta_1=\Upsilon (1S)
-\eta_b(1S) =28 $ MeV found in our calculations appears to be two times
smaller than that in Ref. \cite{ref.10}, where $\Delta_1 =60$ MeV was
obtained.

We consider that the use of the w.f. at the origin $|\tilde
R_n(0)|^2_{\rm exp}$, extracted from  di-electron widths, is the most
promising because these w.f.s take implicitly into account the
relativistic corrections as well as the influence of open channel(s),
in this way drastically simplifying the theoretical analysis. A
comparison of these w.f.s with those calculated in different models
puts serious restrictions on the static potential used and also on
many-channel models.

We also show that the nonperturbative spin-spin interaction gives a
contribution of about 9 MeV in the $J/\psi-\eta_c (1S)$ mass
difference.

\section{The spin-spin interaction}

The HF splitting between the $n^3S_1$ and $n^1S_0$ levels will be
considered here for two cases. The first one corresponds to the
standard perturbative (P) spin-spin interaction with a
$\delta$-function:
\begin{equation}
 \hat V^{\rm P}_{ss} (r) = \ves_1\cdot\ves_2\,
 \frac{32\pi}{9\omega^2_q} \alpha_s (\tilde \mu)
 \left(1+\frac{\alpha_s}{\pi}\rho\right) \delta (\ver)
 \equiv\ves_1\cdot \ves_2 V_{\rm HF}(r),
\label{1a}
\end{equation}
which in one-loop approximation gives the following HF splitting \cite{ref.06}:

\begin{equation} \Delta^{\rm P}_{\rm HF}(nS) =
 \frac{8}{9} \frac{\alpha_s(\tilde \mu)}{\omega^2_q} |R_n(0)|^2
 \left(1+\frac{\alpha_s(\tilde \mu)}{\pi}\rho \right).
\label{2}
\end{equation}  
where $\rho=\frac{5}{12} \beta_0 -\frac83-\frac34 \ln 2$.  This
correction is small: $ \la 0.5\%$ in bottomonium $(n_f=5)$ and $\la 3\%$ 
in charmonium $(n_f = 4)$.

It is very probable that $\delta(\ver)$ may be considered as a limiting
case and the ``physical'' spin-spin interaction is smeared with a still
unknown ``smearing'' radius. For example, for the Gaussian smearing
function,
\begin{equation}
 \delta(\ver) \to \frac{4\beta^3}{\sqrt{\pi}}
 \int r^2 dr \exp (-\beta^2 r^2),
\label{4}
\end{equation}
the splitting can be rewritten as
\begin{equation}
\Delta^{\rm P}_{\rm HF} (nS) =\frac{8}{9}
\frac{\alpha_s(\tilde \mu)}{\omega^2_q}\xi_n(\beta)
|R_n(0)|^2 \left( 1+\frac{\alpha_s}{\pi}\rho\right ),
\label{5}
\end{equation}
where by definition ``the  smearing factor" $\xi_n (\beta)$ is
\begin{equation}
 \xi_n (\beta) =\frac{4}{\sqrt{\pi}}
 \frac{\beta^3}{|R_n(0)|^2} \int |R_n(r)|^2 \exp (-\beta^2r^2) r^2 dr.
\label{6}
\end{equation}
The general  expression Eq.~(\ref{5}) is evidently valid for other smearing
prescriptions which may differ from Eq.~(\ref{4}). The factor $\xi_n$ is
calculated in Appendix~\ref{appendix.A}.

It is well known \cite{ref.08} that the w.f. at the origin is very
sensitive to the form and parameters of the gluon-exchange interaction
and also to the value of the quark mass used. Therefore we make the
following remarks
\begin{enumerate}
\item  To minimize the uncertainties in the w.f. at the origin,
$R_n(0)$,  we shall use here the w.f.s extracted from the experimental
data on the leptonic widths and denote them as $|\tilde R_n(0)|^2_{\rm
exp}$. In this way the relativistic corrections to the w.f. and the
influence of open channel(s) are implicitly taken into account.

\item
In Eqs.~(\ref{2}) and (\ref{5}) the constituent mass $\omega_q$  enters.
This fact can be rigorously deduced from relativistic calculations
\cite{ref.14,ref.15},
\begin{equation}
 \omega_q (nS) =\langle \sqrt{\vep^2 + m^2_q}\rangle_{nS},
\label{7}
\end{equation}
where under the square-root the pole mass $m_q\equiv m_q$(pole) is
present. This mass is known with good accuracy and we take here
$m_b({\rm pole}) =4.8 \pm 0.1$ GeV and $m_c({\rm pole}) =1.42\pm 0.03$
GeV, which correspond to well-established current masses $\bar m_b
(\bar m_b)=4.3(1)$ GeV, $\bar m_c(\bar m_c) =1.2(1)$ GeV
\cite{ref.03}.  The important feature of the constituent mass
$\omega_q(nS)$ is that it takes into account the relativistic
correction and its values depends  weakly on the quantum numbers and
on the static interaction used. Indeed, they lie in a rather narrow range
for all $nS$ states, both in charmonium and bottomonium:
\begin{eqnarray}
 \omega_c(1S)&  = & 1.62 \pm 0.03~ {\rm GeV} (n=1),
 \quad \omega_c(nS)=1.71 \pm 0.03~{\rm GeV} (n\geq 2),
\nonumber \\
 \omega_b(nS) & = & 5.05 \pm 0.15~{\rm GeV}.
\label{8}
\end{eqnarray}
Note that just these mass values are mostly used in nonrelativistic
calculations, thus  in a hidden way taking into account relativistic
corrections.

\item
The leptonic width of the $n^3S_1$ state  in heavy quarkonia is
defined  by the Van Royen-Weisskopf formula with QCD correction
$\gamma_q$ \cite{ref.16a}:
\begin{equation}
 \Gamma_{ee} (n^3S_1)|_{\rm exp} =
 \frac{4e^2_q\alpha^2}{M^2_n} |\tilde R_n(0)|^2_{\rm exp} \gamma_q.
\label{a8}
\end{equation}
Here $e_q=\frac{1}{3} \left( \frac{2}{3}\right)$ for a $b(c)$ quark;
$\alpha=1/137$, $M_n\equiv M(n^3S_1)$, and the QCD factor is
given by
\begin{equation}
 \gamma_q(nS) = 1-\frac{16}{3\pi}\alpha_s(2m_ q).
\label{9} 
\end{equation}

\end{enumerate}

The renormalization scale $\mu$ in Eq.~(\ref{9}) is supposed to be
known, $\mu=2m_q$(pole), as in Refs.~\cite{ref.09,ref.10} and in
$\eta_b \to \gamma\gamma$ decay \cite{ref.16}. In some cases $\mu=M_n$
is also taken.  With an accuracy of $\la 1\%$ in bottomonium and
$\la 10\%$ in charmonium both choices coincide and therefore one can
take here $\mu=2m_q$ $(2m_b=9.6$ GeV, $2m_c=2.9$ GeV).

Since for $n_f=5$ the QCD constant $\Lambda^{(5)}_{ \overline{MS}}$ is
well known from  high energy experiments \cite{ref.03}, the factor
$\gamma_b$ is supposed to be determined with a good accuracy. For
$\Lambda^{(5)}_{ \overline{MS}}(3-{\rm loop}) =210(10)$ MeV, which
corresponds to $\alpha_{ \overline{MS}}(M_Z)=0.1185$, one finds
\begin{equation}
 \gamma_{b}=\gamma_{bn} =0.700(5), \quad \alpha_s(2m_b) =0.177(3).
\label{10}
\end{equation}
In charmonium $(n_f=4)$ the strong coupling $\alpha_{ \overline{MS}}
(2m_c=2.9$ GeV) is determined with less accuracy and for $\Lambda^{(4)}_{
\overline{MS}} =0.260(10)$ MeV one obtains
\begin{equation}
 \alpha_s (2m_c=2.9 ~{\rm GeV})~=0.237(5), \quad \gamma_c =0.60(2).
\label{11}
\end{equation}
Here, the theoretical error comes from the uncertainty in the value
$\Lambda^{(4)}_{ \overline{MS}}.$ Then the w.f. at the  origin,
extracted from the di-electron width  in Eq.~(\ref{8}),
\begin{equation}
 |\tilde R_n(0)|^2_{\rm exp} =\frac{M^2_n\Gamma_{ee}
 (n^3S_1)}{4e^2_q \alpha^2\gamma_{q}},
\label{12}
\end{equation}
implicitly takes into account the relativistic corrections as well as
the influence of the open channels, which gives rise to smaller values
for $|R_n(0)|$ and the HF splitting.  An additional decrease of
$\Delta_{\rm HF}$ comes from a possible smearing of the $S$-wave
spin-spin interaction (for the $P$-wave states this effect is very
small).

The extracted values of $|\tilde R_n(0)|^2_{\rm exp}$ in the $b\bar b$
and $c\bar c$ systems are presented in Tables~\ref{table.1} and
\ref{table.2}.

With the use of these w.f.s at the origin and the values $\gamma_b
=0.70$ we also very precisely reproduce the ratios of the leptonic
widths measured by the CLEO Collaboration (third Ref. in
\cite{ref.12}):
\begin{eqnarray}
 \Gamma_{ee} (\Upsilon(2S))/\Gamma_{ee} (\Upsilon(1S))& =& 0.46 (1),
 \quad \Gamma_{ee} (\Upsilon(3S))/\Gamma_{ee} (\Upsilon(1S)) = 0.32(1)
\nonumber \\
 \Gamma_{ee} (\Upsilon(3S))/\Gamma_{ee} (\Upsilon(2S))& =& 0.69 (2),
\end{eqnarray}
which confirms our correct choice of equal values of $\gamma_{bn}$ for
these states.

\begin{table}
\begin{center}
\caption{\label{table.1}
The w.f. $|\tilde R_n(0)|^2_{\rm exp}$ from Eq.~(\ref{12})
(in GeV$^3$) and the leptonic widths $\Gamma_{ee} (\Upsilon(nS))$
(in keV) for the $\Upsilon(nS)$ states ($\gamma_b=0.70)$ ${}^{a)}$}
\vspace{1ex}
\begin{tabular}{|c|c|c|}
\hline
&&\\
 &~$\Gamma_{ee} (nS)_{\rm exp}$& $|\tilde R_n(0)|^2_{\rm exp}~^{b)}$\\
&&\\
\hline
&&\\
&$1.314(29)$&7.094(16)\\
1S&&\\
&1.336(28) &7.213(15)\\
&&\\\hline &&\\
&0.576(24) & 3.49(15)\\2S&&\\
&0.616(19) & 3.73(12)\\
&&\\ 
\hline
&&\\
3S&0.413(10) & 2.67(7)\\
&&\\
\hline
&&\\
4S&0.25(3) &1.69(20)\\
&&\\
\hline
&&\\
5S& 0.31(7) &2.21(49)\\
&&\\
\hline 
&&\\
6S&0.13(3) &0.95(22)\\ 
&&\\
\hline

\end{tabular}

\end{center}

$^{a)}$ For the states $1S$ and $2S$ the upper entries are taken from
PDG \cite{ref.03} and the lower ones from the CLEO data \cite{ref.12}.
The numbers concerning the $3S$ state are also taken from the CLEO data
\cite{ref.12}. The values for the states $4S$, $5S$, and $6S$ are
taken from PDG \cite{ref.03}.

$^{b)}$ In the third  column only experimental errors are  given.\\

\end{table}

It is of interest to compare the extracted values of $|\tilde
R_n(0)|^2_{\rm exp}$ with the theoretically predicted values, which
mostly depend on the  strong coupling used in the gluon-exchange (GE)
term. In particular, if the asymptotic freedom (AF) behavior of
$\alpha_{\rm static} (r)$ is neglected, then the theoretical values can
be $2 - 1.5$ times larger than $|\tilde R_n(0)|^2_{\rm exp}$, even for
those  $\Upsilon (nS)(n=1,2,3)$ states, which lie far below the $B\bar
B$ threshold \cite{ref.08} (see Table~\ref{table.6} in
Appendix~\ref{appendix.A}).

\begin{table}

\begin{center}

\caption{\label{table.2} The w.f. $|\tilde R_n(0)|^2_{\rm exp}$ (in GeV$^3$)
and the leptonic widths $\Gamma_{ee} (n~^3S_1)$ (in keV ) in
charmonium ($\gamma_c =0.60)$${}^{a)}$.}
\vspace{1ex}
\begin{tabular}{|c|c|c|c|c|}\hline
&&&&\\
&1S&2S&3S&4S\\ 
&&&&\\
\hline 
&&&&\\
&5.40(22)&2.12(12) & 0.75(1) & 0.47(15)) \\
$\Gamma_{ee}({\rm exp})$&&&&\\
&5.68(24)& 2.54(14) & 0.89(8) & 0.71(10)\\
&&&&\\
\hline
&&&&\\
&0.911(37)&0.51(3)&$ 0.22(1)$&
$0.16(5)$\\
$|\tilde R_n(0)|^2_{\rm exp}$&&&&\\
&0.959(40)&0.61(3)&$ 0.26(2)$&
$0.24(4)$\\
&&&&\\
\hline

\end{tabular}

\end{center}

$^{a)}$ The upper entries are taken from Ref.~\cite{ref.03}
and the lower ones are taken from Ref.~\cite{ref.13}.

\end{table}

In our previous analysis of the spectra and fine structure splittings in
heavy quarkonia \cite{ref.07,ref.14,ref.15} we have used the static
potential $V_B(r)$ in  which  the strong coupling in coordinate space
$\alpha_B(r)$ is defined as in background perturbation theory (see
Eq.~(\ref{A.3})). In bottomonium (in the single-channel approximation) the
potential $V_B(r)$  appears to give values of
$|R_n(0)|^2_{\rm th}$  very close to the numbers $|\tilde
R_n(0)|^2_{\rm exp}$.  For illustration in Table~\ref{table.3}  the ratios
\begin{equation}
 S_n =\frac{|\tilde{R}_n(0)|^2_{\rm exp}}{|\tilde R_n(0)|_{\rm th}^2}
\label{13}
\end{equation}
are given for all known $nS$ levels in charmonium and bottomonium.

\begin{table}
\begin{center}
\caption{\label{table.3} The  factor $S_n$  defined by Eq.~(\ref{13})
for the
potential $V_B(r)$ from Eq.~(\ref{A.2}) for the $\Upsilon(nS)$ and
$\psi(nS)$ wave functions at the origin$^{a)}$.}
\vspace{1ex}
\begin{tabular}{|c|c|c|c|c|c|c|}\hline
&&&&&&\\
 &$1S$&$2S$&$3S$&$4S$&$5S$&$6S$\\ 
&&&&&&\\
\hline 
&&&&&&\\
$b\bar b$ &1.08(4) &1.02(4)&1.02(4)&$ 0.72 (9)$&1.03(22)&0.47(10)\\
&&&&&&\\
\hline
&&&&&&\\
$c\bar c$&1.01(4)&0.82(5)&$0.41(2)$&$0.32(10)$&&\\
&&&&&&\\
\hline

\end{tabular}

\end{center}

$^{a)}$ Given numbers correspond to $\Gamma_{ee}$ taken from PDG
\cite{ref.03}.

\end{table}

As shown in Table~\ref{table.3}, in bottomonium for the potential
$V_B(r)$ the influence of open channels appears to be important only
for the $4S$ and $6S$ levels, while for other states the single-channel
calculations are in good agreement with experiment. It is not so for
many other potentials \cite{ref.08} and this means that any conclusions
about the role of open  channels cannot be separated from the $q\bar q$
interaction used in a given theoretical approach.

In charmonium the effect of open channels is much stronger and reaches
$\sim  60\%$ for the $3S$ and the $4S$ states ($S_n  \cong0.4$) and
about 20\% for the $\psi (2S)$ meson. This number is even smaller (12\%)
if the new CLEO data from Ref.~\cite{ref.13} are used.

\section{The hyperfine splittings in bottomonium}

We consider two cases:

\begin{description}
\item[A.] No  smearing  effect, i.e. in Eq.~(\ref{5}) the
smearing parameter $\xi_{bn} =1.0.$ (for all $n$).
\item[B.] The smearing parameter $\xi_{bn}$, Eq.~(\ref{6}), is
calculated with the value $\beta=\sqrt{12}$ GeV, corresponding to the
smearing radius $r_{ss}=\beta^{-1} =0.057$ fm.  
\end{description}

Unfortunately, at present there is no precise prescription as to how to
choose the renormalization scale in the HF splitting Eq.~(\ref{2}): in
$\alpha_{\overline{MS}}(\tilde \mu)$ the scale $\tilde \mu=m_b({\rm
pole}) \cong 4.80 \pm 0.01$ GeV is often used. With
$\Lambda_{\overline{MS}}(n_f=5)=210 (10)$ MeV (just the same as in our
calculations of $\gamma_b$ Eq.~(\ref{10})) we find
\begin{equation} 
 \alpha_s(b\bar b,\tilde \mu)=\alpha_{\overline{MS}} (4.8 ~{\rm GeV}) =0.21(1). 
\label{14} 
\end{equation}
With this value of $\alpha_s(\tilde \mu)$ and $|\tilde R_n(0)|^2_{\rm
exp}$ from Table~\ref{table.1}, one obtains the HF splittings in
bottomonium presented in Table~\ref{table.4}, second column.

\begin{table}
\begin{center}
\caption{\label{table.4} $\Delta^P_{\rm HF} (nS) $ (in MeV) in bottomonium for
$\alpha_{\overline{MS}}(M_n) = 0.21; \omega_b=5.10$ GeV and $|\tilde R_n
(0)|^2_{\rm exp}$ from Table~\ref{table.1}.}
\vspace{1ex}
\begin{tabular}{|c|c|c|}
\hline
&&\\
 &~$\xi_b=1.0$~&~$\xi_{bn}$ from Appendix~\ref{appendix.B}\\
&&$\beta=\sqrt{12}$ GeV\\
&& $r_{ss} = 0.057$ fm\\
\hline
&&\\
1S&51 (4) (4)& 28 (2) (3)\\&&\\\hline &&\\2S&25 (3) (2) &12 (2) (1)\\
&&\\ 
\hline
&&\\
3S&22 (5) (2)& 10 (2) (1)\\&&\\\hline
&&\\
4S&12 (3) (1)&5.1 (2) (1)\\
&&\\
\hline
&&\\
5S& 16 (2) (1) &6.4 (1) (1)\\
&&\\
\hline 
&&\\
6S&7 (2) (1)&2.7 (1) (1)\\ &&\\
\hline

\end{tabular}

\end{center}
\end{table}

The numbers in Table~\ref{table.4} contain an experimental error coming
from $\Gamma_{ee} (nS)$ \cite{ref.03} (first number), and a theoretical
error (second number) $\la 10\%$ ( $\la 4\%$ from the $\omega_b$ value
and $\la 5\%$  comes from $\alpha_{\overline{MS} }(m_b)$). For
the $\Upsilon (nS)$ states $(n=1,2,3)$ the calculated  HF splittings
($\xi_n=1.0$) are very close to the splittings from Refs.~\cite{ref.09}.

In Table~\ref{table.4} the HF splittings, calculated with the smearing
function Eq.~(\ref{6}), are also given (third column).  The smearing
radius, $r_{ss}=\beta^{-1}=0.057$ fm ($\beta =\sqrt{12}$ GeV), is taken
to  fit the experimental values of the $J/\psi-\eta_c(1S)$ and
$\psi(2S)-\eta_c(2S)$ splittings (see the next section). However, even
for  such a small radius  $\Delta_{\rm HF}(nS)$ turns out to be
$50\%~(n=1 - 4)$, $60\%~(n=5,6)$ smaller compared to the ``nonsmearing''
case. In particular, the $\Upsilon(1S) -\eta_b(1S)$ splitting turns
out to be  28 MeV instead of 51(4) MeV for  $\xi_{bn} =1.0$. For
higher excitations very small splittings, $\Delta_{\rm HF} \approx 6$
MeV and 3 MeV,  for the $5S$ and $6S$ states are obtained.

Our value of $r_{ss}=0.057$ fm is very close to that from
Ref.~\cite{ref.10} where $r_{ss}=0.060$ fm is taken, both in
charmonium  and in bottomonium. However, in spite of this coincidence
our numbers are about two times smaller than in \cite{ref.10}, where
$\Upsilon(1S)-\eta_b(1S)$ = 60 MeV is obtained. For the $2S$ state our
value of splitting is 12 MeV, still smaller than the value 20 MeV found in
Ref.~\cite{ref.10}.

We can conclude that observation of an $\eta_b(nS)$ meson could
clarify the role of smearing in the spin-spin interaction between a
heavy quark and antiquark.

\section{$\eta_c(nS)$ masses}

The splitting Eq.~(\ref{5}) factually depends on the product
$\alpha_s(\tilde \mu)\times \xi_n$, therefore it is convenient to
discuss an  \underline{effective } HF coupling:
\begin{equation}
 \alpha_{\rm HF} (nS) =\alpha_s (\tilde \mu_n)\, \xi_{cn},
\label{15}
\end{equation}
which  is the only unknown factor since $|\tilde R_n(0)|^2$ can be
extracted from the di-electron widths and the constituent masses
$\omega_c(nS)$ are known with $\la 5\%$ accuracy from relativistic
calculations. These masses  may be specified for different $nS$ states,
Refs.~\cite{ref.14,ref.15}:
\begin{eqnarray}
 \omega_c(1S) & = & 1.62(3) ~{\rm GeV}, \quad \omega_c(2S)=1.71(4) ~{\rm GeV},
\nonumber\\
 \omega_c(3S)\ & \approx  &  \omega_c(4S)=1.73(4) ~{\rm GeV}.
\label{16}
\end{eqnarray}
Here the theoretical errors come from a variation of the parameters of
the static potential.

As discussed in Ref.~\cite{ref.07}, the experimental splittings $J/\psi
-\eta_c (1S)$ and $\psi(2S)-\eta_c (2S)$, can be described if a
different $\alpha_{\rm HF}$ Eq.~(\ref{15}) for the $1S$ and $2S$ states
are taken, namely, $\alpha_{\rm HF} $(1S) $\cong 0.36$ and $\alpha_{\rm
HF} $(2S) $\cong 0.30.$ Such a choice implies two possibilities:
\begin{description}
\item[A] $\alpha_s(\mu_1) =0.36$, $\alpha_s(\mu_2) =0.30$,
$\alpha_s(\mu_3) = \alpha_s(\mu_4)\leq 0.30$,
$\xi_{cn} =1.0 $ (for~all~$n$).
i.e., the renormalization scale is supposed to grow for larger
excitations. In particular, for the value
$\Lambda^{(4)}_{\overline{MS}} (2-{\rm loop})=270$ MeV we have
$\mu_1=1.25$ GeV $\cong\bar m_c(\bar m_c)$, while the scale $\mu_2=1.60
$ GeV is essentially larger. For this choice of $\alpha_{\rm HF}$ the
perturbative HF splittings are  given in Table~\ref{table.5}.

\item[B] The normalization scale $\mu_n$ (and therefore the coupling
constant $\alpha_s(\mu_n)$) is supposed to be equal for all $nS$-states:
$\alpha_s(\mu_n) = 0.36$. In this case a smearing of the spin-spin 
interaction is of principal importance to explain the experimental
value for the $\psi(2S) - \eta_c(2S)$ splitting.
\end{description}

\begin{table}
\begin{center}
\caption{\label{table.5} The splittings $\Delta^{\rm P}_{\rm HF}(nS)$ and 
$\Delta^{\rm NP}_{\rm HF}(nS)$ (in MeV) in charmonium$^{a)}$.}
\vspace{1ex}
\begin{tabular}{|c|c|c|c|}
\hline
&&&\\
 &~$\Delta^{\rm P}_{\rm HF}(nS)$~&~$\Delta^{\rm P}_{\rm HF}(nS)$&
 $\Delta^{\rm NP}_{\rm H F}(nS)^{b)}$\\
&(no smearing: $\xi_c=1.0)$&$r_{ss} =0.29$ GeV$^{-1}$& $G_2=0.043$ GeV$^4$\\
&$\alpha_s(\mu_1)=0.36$&  $\alpha_s(\mu_n)=0.36$& \\
&$\alpha_s(\mu_n) =0.30$&$(n=1 - 4)$  & \\
&$(n=2,3,4)$& &\\
&&&\\
\hline
&&&\\
 $1S$ &118(5)& 100(6) & $9\pm 2$\\
&&108(7)$^{c)}$&\\
&&&\\
\hline 
&&&\\
experiment& 117(2)& 117(2)&\\ 
$J/\psi-\eta_c(1S)$ &&&\\&&&\\
\hline 
&&&\\
$2S$ &51(5)&46(3)&$3.5\pm 1.5$\\
&61(5)$^{c)}$&55(4)$^{c)}$& \\
&&&\\ 
\hline
&&&\\
experiment &48(4)&48(4)&\\   
$\psi(2S)-\eta_c (2S)$ &&&\\
&&&\\
\hline 
&&&\\
 $3S$ &21(2)& 16(2)&$2\pm 1$\\
&&&\\
\hline
&&&\\
 $4S$ &15(4)&12(4)&$1.5\pm 0.5$\\
&&&\\
\hline

\end{tabular}

\end{center}

$^{a)}$ The w.f. $|\tilde R_n(0)|^2_{\rm exp}$ is taken from
Table~\ref{table.2} and corresponds to $\Gamma_{ee}(nS)$ from PDG
\cite{ref.03}.

$^{b)}$ The NP splittings are calculated in  Appendix~\ref{appendix.C}.

$^{c)}$ Here $|\tilde R_1(0)|^2_{\rm exp}=0.959 $ GeV$^3$ and $|\tilde
R_2(0)|^2_{\rm exp} =0.61$GeV$^3$ are extracted from the CLEO data
\cite{ref.13}.

\end{table}

Besides, we have also calculated the contributions coming from the NP
spin-spin interaction.  In bottomonium their values are small,
$\Delta_{\rm HF}^{\rm NP} (nS) < 1$ MeV and can be neglected. In
charmonium, as well as in light mesons, the situation is different. Due
to the NP spin-spin interaction in the $1P\, c\bar c$ state a cancellation
of the perturbative and NP terms takes place \cite{ref.17}. As a result,
the mass difference $M_{\rm cog} (\chi_{cJ}) -M(h_c) =(1\pm 1)$ MeV turns
out to be close to zero or even positive, in accord with 
experiment \cite{ref.18}. Just the same NP contribution, Eq.~(\ref{19}),
provides the correct value of the splitting $M_{\rm cog} (a_J)-M(b_1
(1P))\approx 22$ MeV in light mesons \cite{ref.19}.

The values of $\Delta_{\rm HF}^{\rm NP}(nS)$, (in Table~\ref{table.5},
fourth column) are calculated in Appendix~\ref{appendix.C}. They can be
slightly different for different values of the gluonic correlation
length $T_g$ which defines the behavior of the vacuum correlation
functions (v.c.f.). For  $T_g =0.3$ fm the v.c.f. has an exponential
behavior over the whole range, $0<r<\infty$ \cite{ref.20} and in this
case
\begin{equation}
 \Delta^{\rm NP}_{\rm HF} (nS) =\frac{\pi^2}{18} \frac{G_2}{\omega^2_c}
 1.20 (1\pm 0.07)\mathcal{F} (nS),
\label{19}
\end{equation}
where the number 1.20(1$\pm 0.07)$ follows from lattice data
\cite{ref.20}. In Eq.~(\ref{19}) the gluonic condensate $G_2$ is taken
to be equal to 0.043(3) GeV$^4$, as in Ref.~\cite{ref.17}, and the matrix
element (m.e.) is defined as
\begin{equation}
 \mathcal{F}(nS) = \langle r K_1 \left(r/T_g\right) \rangle_{nS}.
\label{20}
\end{equation}
Our calculations give
\begin{eqnarray}
 \mathcal{F}(nS)  = & & 0.80 {\rm ~GeV}^{-1} (1S),~~0.40 {\rm ~GeV}^{-1} (2S),
\nonumber \\
 & & 0.27 {\rm ~GeV}^{-1} (3S),~~ 0.20 {\rm ~GeV}^{-1}(4S).
\label{21}
\end{eqnarray}
The corresponding $\Delta_{\rm HF}^{\rm NP} (nS)$ are given in
Table~\ref{table.5}. One can see that $\Delta_{\rm HF}^{\rm NP} (1S)
=9\pm 2$ MeV needs to be taken into account in the $J/\psi-\eta_c(1S)$
splitting while for the higher states the values of $\Delta^{\rm
NP}_{\rm HF}$ lie within the experimental and theoretical errors.
Their values are $\Delta_{\rm HF}^{\rm NP} =$ 3(2) MeV, 2(1) MeV, and
1.5(5) MeV respectively for the $2S$, $3S$, and $4S$ states. A smaller
value of $T_g (T_g \la 0.2 $ fm), which also cannot be excluded,
practically does not change the numbers. Adoption of this value would
only slightly decrease those splittings. Thus one can conclude that in
case {\bf A}  with different renormalization scales $\mu_n (\mu_1 \approx
1.25 $ GeV is small, $\mu_2=1.6$ GeV),  for the $J/\psi-\eta_c (1S)$
and $\psi$ (2S) $-\eta_c (2S)$ splittings agreement with experimental
values can easily be obtained.

If the renormalization scales $\mu_n$  are  supposed to be equal for
all $nS$ states:  $\alpha_s( \mu_n\cong\bar m_c= 1.25$ GeV ) =0.36,
then to explain the relatively small  $\psi (2S) - \eta_c(2S)$
splitting a smearing effect needs to be introduced. Then for the
potential used the values $\xi_n (c \bar c)= 0.85,~0.80,~0.78,~0.76$
for the $1S$, $2S$, $3S$, and $4S$ states respectively, are calculated
in Appendix~\ref{appendix.B} and for this case the values of
$\Delta_{\rm HF}^{\rm P} (c\bar c, nS)$ are also given in
Table~\ref{table.5}. For the $3S$ and $4S$ levels the values we predict
are about 21(15) MeV  (no smearing) and 16(12) MeV (with smearing),
i.e. the  difference between the cases {\bf A} and {\bf B} is only $\sim 20\%$.
Notice that in case {\bf B} the NP contribution improves the agreement
with experiment for the states $J/\psi-\eta_c(1S)$.

Thus, in the $\psi(nS) -\eta_c (nS)$ splittings the smearing effect
appears to be less prominent that in bottomonium.

In Appendix~\ref{appendix.B} we also show that the relativistic
correlations as well as the constituent masses stop to grow when
the many-channel description is effectively used. This fact is very
important for the study of higher excitations in charmonium.

\section{Conclusions}

In our paper we have shown that
\begin{enumerate}
\item In bottomonium $\Delta_{\rm HF}^{\rm P} (nS)$ appears to be very
sensitive to smearing of the spin-spin interaction. Due to this effect
the splitting decreases  from 51 MeV  to 28 MeV for the $1S$ state and
from 25 Mev  to 12 MeV for the $2S$ state; very small values are
obtained for higher states.
\item  In charmonium there are two possibilities to describe
$\Delta_{\rm HF} (1S)$ and $\Delta_{\rm HF} (2S)$, which are  known
from experiment. The first one refers to a different choice of the
renormalization scale: $\mu_1$ =1.25 GeV and $\mu_2\cong 1.60$ GeV for
the $1S$ and $2S$ states, if the smearing effect is absent. The second
possibility implies equal renormalization scales $\mu_n (n=1 - 4)$
for all $nS$ states. Then to explain the $\psi (2S) -\eta_c (2S)$
splitting the smearing of the spin-spin interaction needs to be taken
into account. We also  expect that for the $1S$ level a small contribution
($\sim $ 9 MeV) comes from the NP spin-spin interaction.
\item The $\psi (3S)- \eta_c (3S)$ splitting is predicted to be
around 16(2) MeV, without and  12(4) MeV with smearing effect.
\end{enumerate}
To understand the true role of the smearing effect in the spin-spin
interaction the observation of an $\eta_b(nS)$ is crucially important.

\appendix

\section{The wave function at the origin}
\label{appendix.A}
\setcounter{equation}{0} \def\theequation{A.\arabic{equation}}

Two factors strongly affect the w.f. at the origin:
\begin{description}
\item[(i)] The AF behavior of the coupling in the GE term: 
$ V_0(r) =\sigma r -\frac{4}{3} \frac{\alpha_{GE}(r)}{r}$,

\item[(ii)]  The influence of open channels.
\end{description}
For illustrations of these statements in Table~\ref{table.6} the w.f.s
$|R_n(0)|^2_{\rm th}$ are given for several cases.
\begin{enumerate}
\item For the Cornell potential, $V_c(r) =\sigma r -\frac43
\frac{\alpha_c}{r}$ in single-channel approximation ($\alpha_c=0.39,
\sigma=(2.34)^{-2}$ GeV$^2, \omega_c=1.84$ GeV).
\item For the same  Cornell potential while open channels are taken
into account  with the use of the nonrelativistic Cornell
coupled-channel (CCC) model \cite{ref.21}. Then the w.f. in the
coupled-channel system, $\psi_n(r) \equiv \psi(n^3S_1)$,  is
presented as the composition of different  contributions, e.g.

\begin{eqnarray}
 \psi_1(1^3S_1)&=&0.983|1S>-0.050|2S> -0.009|3S>-0.003|4S>,
\nonumber\\
 \psi_2(2^3S_1)&=&0.103|1S>+0.838|2S> -0.085|3S>
\nonumber\\
 & & \hspace{-0.8em} -0.017|4S>-0.007|5S>+0.040|1D>-0.008|2D>,
\nonumber\\
 \psi_3(3^3S_1)&=&0.02 e^{-i0.05\pi}|1S>+0.19 e^{-i0.30\pi}|(2S) >+ 0.67|3S>
\nonumber\\
 & & \hspace{-0.8em} +0.07 e^{i0.54\pi} |4S > + 0.04 e^{i0.59\pi}
 |1D>+0.04e^{i0.59\pi}|2D>.
\nonumber \\
\label{A.1} 
\end{eqnarray}
In single-channel approximation the w.f. $|R^{\rm C}_{nS} (0)|^2$  for the
Cornell potential and the derivatives $|R^{\prime\prime}_D(0)|^2$ were
calculated in \cite{ref.08}:  $|R^{\rm C}_{1S} (0)|^2=1.454$ GeV$^3$,
$|R^{\rm C}_{2S} (0)|^2=0.927$ GeV$^3$, $|R^{\rm C}_{3S} (0)|^2=0.791$ GeV$^3$,
$|R^{\rm C}_{4S} (0)|^2=0.725$ GeV$^3$; $|R^{\prime\prime}_{1D}
(0)|^2_{\rm C}=0.030$ GeV$^7$;  $|R^{\prime\prime}_{2D} (0)|^2_{\rm C}=0.0655$
GeV$^7$. From Eq.~(\ref{A.1}) it follows that in the CCC model the $2^3S_1
-1^3D_1$ and the $3^3S_1-2^3D_1$ mixings appear to be small compared to
the analysis in \cite{ref.22}, where the admixture of the $1^3D_1$
state is $\sim 22\%$.

\item
For the potential

\begin{equation} V_B(r) =\sigma r -\frac43 \frac{\alpha_B(r)}{r},
\label{A.2}
\end{equation}
the coupling
is defined  as in Refs.~\cite{ref.14,ref.15}:

\begin{equation} \alpha_B(r) =\frac{8}{\beta_0}\int
dq\frac{\sin qr}{q} \frac{1}{t_B(q)} \left[
1-\frac{\beta_1}{\beta_0^2} \frac{\ln
t_B}{t_B}\right]
\label{A.3}
\end{equation}
with
$$t_B(q) =\ln \frac{q^2+M^2_B}{\Lambda^2_B(n_f)}.$$

\end{enumerate}
Here $M_B =0.95$ GeV, is the background mass,  $\Lambda_B (n_f)$ is
expressed through the $\Lambda_{\overline{MS} }(n_f)$ and in 2-loop
approximation $\Lambda_B (n_f =4) =0.360(10)$ MeV \cite{ref.14}. The
string tension $\sigma=0.18$ GeV$^2$ and  the same values of $\omega_c
=1.65$ GeV are taken here for simplicity for all $nS$ states.

In Table~\ref{table.6} for comparison the values of $|\tilde
R_n(0)|^2_{\rm exp}$ extracted from the di-electron widths are also
given.  Then  one can see that in single-channel approximation  very
different  $ | R_n(0)|^2_{\rm th}$ are obtained for $V_c(r)$ and $V_B(r)$.
For the potential  $V_B(r)$ the w.f. $ |R_1(0)|^2$  appears to be very
close to the ``experimental'' number: $|\tilde R_1(0)|^2=0.91$ for $J/\psi$,
while for the Cornell potential $|R_n(0)|^2_{\rm th}$ is about two times
larger than $|\tilde R_n(0)|^2_{\rm exp}$ (single-channel
approximation). Moreover, even in the CCC model $| R_n(0)|^2_{\rm th}$ is
still too large, being $\sim 15\% (30\%)$ larger for the $2S(1S)$
states compared to $|\tilde R_n(0)|^2_{\rm exp}$. Correspondingly,
for such multi-channel w.f.s the leptonic widths will be 15\% and 30\%
larger than the experimental value, respectively.

\begin{table}
\begin{center}
\caption{\label{table.6} Comparison  of $| \tilde R_n(0)|^2_{\rm exp}$ (in
GeV$^3$) from Eq.~(\ref{12}) with the w.f. $| R_n(0)|^2_{\rm th}$  for  the
Cornell  potential (single-channel approximation and also  in the
coupled-channel model) and for $V_B(r)$ Eq.~(\ref{A.2}).}
\vspace{1ex}
\begin{tabular}{|c|c|c|c|c|}\hline
&&&&\\
&&1S&2S&3S\\ 
&&&&\\
\hline 
&&&&\\
&single-channel&1.454&0.928&0.790\\
&approximation  &&&\\
Cornell&&&&\\
\cline{2-5}
potential &&&& \\
&coupled-channel&1.268&0.703&0.520\\
&model [21]&&&\\ 
&&&&\\
\hline \multicolumn{2}{|c|}{~}&&&\\
\multicolumn{2}{|c|}{$V_B(r)$,} &0.900&0.616&0.534\\
\multicolumn{2}{|c|}{single-channel approx.}&&&\\
\multicolumn{2}{|c|}{~}&&&\\
\hline\multicolumn{2}{|c|}{~}&&&\\
\multicolumn{2}{|c|}{$|\tilde R_n(0)|^2_{\rm exp}$ from Table $2^{a)}$}&
0.91(4)$_{\rm exp}(5)_{\rm th}$&0.51(3)$_{\rm exp}(2)_{\rm th}$&
0.22(3)$_{\rm exp}(1)_{\rm th}$\\
\multicolumn{2}{|c|}{}&$0.96(4)_{\rm exp} (5)_{\rm th}$&$0.61 (3)_{\rm exp} (2)_{\rm th}$&0.26(3)$_{\rm exp}(1)_{\rm th}$\\
\multicolumn{2}{|c|}{~}&&&\\
\hline

\end{tabular}

\end{center}

${}^{a)}$ See the footnote to Table~\ref{table.2} for the origin of
these numbers.

\end{table}

At this point we would like to  stress that one cannot fit the leptonic
widths $\Gamma_{ee} (nS)$ taking a much smaller value for the factor
$\gamma_c$, Eq.~(\ref{10}) (or larger $\alpha_s(\mu)$ in $\gamma_c$).
Such a change in the renormalization scale would come in contradiction
with the description of other ``annihilation'' decays like
$\eta_c(nS)\to \gamma\gamma$, where just the value
$\alpha_s(\mu=2m_c)=0.24(1)$, as in our analysis is taken.  Also we
would like to notice that in the CCC model the $2^3S_1- 1^3D_1$ and
$3^3S_1-2^3D_1$ mixings are small, as seen from Eqs.~(\ref{A.1}), and
instead large admixtures to the $2S(3S)$ states come from the
neighboring $1S(2S)$ states.

For the $V_B(r)$ potential Eq.~(\ref{A.2}) we have obtained good
agreement with the ``experimental'' $|\tilde R_1(0)|^2_{\rm exp}$ (for
the $J/\psi$ $\sim 5\%$ accuracy) and 20\% discrepancy for $\psi(2S)$.
For higher states  open channels give rise to a suppression of $|\tilde
R_n(0)|^2_{\rm th}$ (single-channel approximation) by $\sim 20\%$ for
$\psi(2S)$ and about 60\% for the $\psi(4040)$ and $\psi(4415)$ mesons.
It is precisely the effect of the open channels that appears to be
responsible for the screening of the GE interaction which was discussed
in Ref.~\cite{ref.23}, where open channels have been considered
effectively through flattening of the confining potential and switching
off of the GE interaction.

\section{The smearing factor $\xi_n$ }
\label{appendix.B}
\setcounter{equation}{0} \def\theequation{B.\arabic{equation}}

For the smearing function Eq.~(\ref{4}) the factor
\begin{equation}
 \xi_n \equiv \xi (nS) =\frac{4\beta^3}{\sqrt{\pi}} \frac{J_n}{|R_n(0)|^2},
\label{eq.B.1}
\end{equation}
is calculated here  for the potential $V_B(r)$ Eq.~(\ref{A.2}). This
factor appears to be  weakly dependent on a variation of the mass
$m_q$. In Table~\ref{table.7} its values are given for the parameter
$\beta=\sqrt{12}$ GeV which corresponds to the smearing radius $r_{ss}
=\beta^{-1}=0.057$ GeV. For smaller $\beta$ (larger $r_{ss}$) the
smearing effect is becoming even larger (the $\xi_n $ are smaller).

\begin{table}
\begin{center}
\caption{\label{table.7} The smearing factor $\xi_n$ Eq.~(\ref{6}) in
bottomonium and charmonium for the potential $V_B(r)$ Eq.~(\ref{A.2}).
$ ~\sigma=0.1826$ GeV$^2$, $\alpha_B(r)$ is defined by Eq.~(\ref{A.3}).}
\vspace{1ex}
\begin{tabular}{|c|c|c|c|}
\hline
&&&\\
 &~$\xi_n(b\bar b)^{a)}$~&~$\xi_{n}(c\bar c)^{b)}$&~$\xi_{n}(c\bar c)^{b)}$\\
&&$\beta=\sqrt{12}$ GeV&$\beta=2$ GeV\\
&&&\\
\hline
&&&\\
1S&0.61&0.85&0.69\\
&&&\\
\hline &&&\\
2S&0.55&0.80&0.62\\
&&&\\ 
\hline
&&&\\
3S&0.51&0.78&0.57\\
&&&\\
\hline
&&&\\
4S&0.49&0.76&053\\
&&&\\
\hline
&&&\\
5S& 0.47&0.74&0.50\\
&&&\\
\hline
\end{tabular}

\end{center}

$^{a)}$ $m_b=5.1$ GeV

$^{b)}$ $\omega_c =1.70$ GeV

\end{table}

From Table~\ref{table.7} one can conclude that in bottomonium (due to
smearing) the $\rm HF$ splittings decrease by a factor of two for all
$nS$ levels ($n=1 - 6)$, while in charmonium the smearing effect is
weaker and mainly important for the $3S$ and $4S$ states (see
Tables~\ref{table.4} and \ref{table.5}).

We would also like to make several remarks about relativistic
corrections.  Partly this correction is taken into account through the
use of the constituent mass $\omega_q$ -- the average over the quark
kinetic energy Eq.~(\ref{7}).
Actually, in the Hamiltonian, or the spinless Salpeter equation (SSE):
\begin{equation}
 (T+V)\psi^R_{nL} (r_1) = M_0 (nL) \psi^R_{nL} (r)
\label{B.2}
\end{equation}
we use the expansion of the square root near the point $\vep^2 + m^2_q
-\omega_q^2$. Then  the kinetic term is
\begin{equation}
 T_R = 2\sqrt{\vep^2+m^2_q}\equiv
 2\omega\sqrt{1+\frac{\vep^2+m^2_q-\omega^2_q}{\omega^2_q}}
 \cong\omega_q +\frac{m^2_q}{\omega_q} +\frac{\vep^2}
 {\omega_q} \equiv T_{EA},
\label{B.3}
\end{equation}
which is different from the standard nonrelativistic case.  This
expansion, or so-called Einbein approximation (EA), appears to be a
good approximation to the SSE \cite{ref.24}. While relativistic
corrections are  important, as in charmonium, it provides much better
accuracy than the NR approximation, where the difference between the
constituent mass $\omega_q (nL)$ and $m_q = $const. is neglected.
Still, for the w.f. at the origin a small difference between the exact
solution $R^R_{nL}(r)$ of the SSE and an approximate solution for the
EA equation occurs
\begin{equation}
 (T_{EA} +V) \psi_{EA} (r) = M_{EA} (nL) \psi_{EA} (r).
\label{B.4}
\end{equation}
As an example, in Table~\ref{table.8} the ratio $\eta=|R^{\rm R}_{nS}
(0)|^2/|R^{\rm EA}_{nS}(0)|^2$ are given for two confining potentials. The first
one refers to the single-channel case with the string tension $\sigma_0= $
const. In this case the relativistic correction (the factor $\eta$) is
growing for higher levels and reaches about 30\% for the $4S$ state.

\begin{table}
\begin{center}
\caption{\label{table.8} The ratio 
$\eta=|R^{\rm R}_{nS} (0)|^2/|R^{\rm EA}_{nS}(0)|^2$ in
charmonium for the linear potential $\sigma_0 r$ and for the flattening
potential $\sigma(r)\cdot r~(\sigma_0=0.18$ GeV$^2$). The mass
$m_c$(pole) $=1.45$ GeV.
The parameters of $\sigma(r) =\sigma_0 [1-\gamma(r)]$ are taken
from Ref.~\cite{ref.23}.}
\vspace{1ex}
\begin{tabular}{|l|c|c|c|c|}
\hline
&&&&\\
 &~$1S$~&~$2S$~&~$3S$~&~$4S$~\\ 
&&&&\\
\hline 
&&&&\\
 $\sigma_0 r$ potential& 1.108& 1.196& 1.258& 1.310\\
&&&&\\
\hline
&&&&\\
 $\sigma (r)\cdot r$ potential & 1.093&1.153&1.147&1.045\\
&& &&\\
\hline

\end{tabular}

\end{center}

\end{table}

However, if there are open channels, the linear potential is
flattening, and such a flattening potential $\sigma(r) \cdot r$
effectively takes into account the open channels.  This feature
provides an essential shift down for high excitations \cite{ref.23a}.
Due to the flattening, already for the $3S$ state the factors $\eta$ as
well as $\omega_c(3S)$, stop to grow, i.e., an open channel affects the
w.f. at the origin  much more strongly than the relativistic
correction. Since at present there is no relativistic multichannel
theory and one cannot distinguish between both effects, we expect that
rather small relativistic corrections ($\leq 15\%)$ will take place for
all $nS$ states in charmonium. Note that this relativistic correction 
does not affect the smearing factor $\xi_n(c\bar{c})$.

\section{Nonperturbative spin-spin interaction }
\label{appendix.C}
\setcounter{equation}{0} \def\theequation{C.\arabic{equation}}

The NP contribution to the HF splitting $M_{\rm cog} (\chi_{cJ} ) - M(h_c)$
has been discussed in detail in Refs.~\cite{ref.17,ref.19}, where the
following general expression has been obtained  
\begin{equation}
 \Delta_{\rm HF}^{\rm NP}(1P) =\frac{\pi^2}{18}
 \frac{g_2}{\omega^2_c} 1.20 (1\pm 0.07) e^{\frac{r_0}{T_g}}\mathcal{F}(nL).
\label{C.1} 
\end{equation}
Here the matrix element is given by
\begin{equation}
 \mathcal{F}(nL) =
 \langle \theta (r-r_0)\,r\, K_1\left(r/T_g\right) \rangle_{nL},
\label{C.2} 
\end{equation}
with $T_g$ the gluonic correlation length, which is still not
rigorously fixed in a lattice measurement (see the discussion in
Ref.~\cite{ref.19}).  The parameter $r_0$ in Eq.~(\ref{C.1})
characterizes the size of the region near $r=0$ where the vacuum
correlation functions are not well defined. In lattice data for large
$T_g=0.3$ fm $(n_f=4)$ it is supposed that $r_0=0$ and then the
expressions (\ref{C.1}) and (\ref{C.2}) go over into
Eqs.~(\ref{19},\ref{20}). If $r_0\neq 0$, (e.g. for $T_g=0.2$ fm when
the value $r_0\cong T_g =0.2$ fm) an  additional term needs to be added
to $\Delta_{\rm HF}^{\rm NP}$ (\ref{C.1}).

At small $r$, $(r\to 0)$, the NP spin-spin potential was defined in
Ref.~\cite{ref.17},
\begin{equation} 
 \hat V^{\rm NP}_{\rm HF} =\ves_1\cdot \ves_2 V^{\rm NP}_{\rm HF},
 \quad V_{\rm HF}^{\rm NP} (r\to 0) =\frac{\pi^2}{9\omega^2_c} G_2 (r_0 +T_g).
\label{C.3}
\end{equation}
Then assuming that the v.c.f. has a kind of plateau near the origin,
one obtains the following contribution from the region $r\leq r_0$
(Eq.~(\ref{16}) in Ref.~\cite{ref.17}:
\begin{equation}
 V_{\rm HF}^{\rm NP} (r\to 0) =\frac{\pi^2}{9\omega^2_c} G_2 (r_0 +T_g)
 \sqrt{ r^2_0-r^2} \,\theta(r_0-r).
\label{C.4}
\end{equation}
Therefore, in the most general case the total NP contribution is
\begin{eqnarray}
 \Delta^{\rm NP}_{\rm HF} (nL) & = & \frac{\pi^2G_2}{18 \omega^2_c}
 \left\{ 1.20 (1\pm 0.07) e^{\frac{r_0}{T_g}} \mathcal{F}(nL) \theta(r-r_0)
 \right.
\nonumber \\
& & \hspace{3em} \left.
 +2(r_0+T_g) \langle\sqrt{r^2_0-r^2}\rangle_{nL}\theta(r_0-r)\right\}.
\label{C.5}
\end{eqnarray}
Note that for $T_g=r_0=0.2$ fm the second term in Eq.~(\ref{C.5}) is
important  only for the $S$-wave states, being small for the $P$-wave
states, i.e. for the splitting $M_{\rm cog} (\chi_{cJ})-M(h_c)$
considered in Ref.~\cite{ref.17}.

\end{document}